\documentclass[conference]{IEEEtran}
\usepackage{amsmath,amssymb,amsfonts}
\usepackage{algorithmic}
\usepackage{graphicx}
\usepackage{textcomp}
\usepackage[table,xcdraw]{xcolor}
\usepackage{url}
\usepackage{hyperref}
\usepackage{todonotes}
\usepackage{float}
\usepackage{multirow}
\usepackage[many]{tcolorbox}    	
\usepackage[shortlabels]{enumitem}   
\def\BibTeX{{\rm B\kern-.05em{\sc i\kern-.025em b}\kern-.08em
    T\kern-.1667em\lower.7ex\hbox{E}\kern-.125emX}}
\ifCLASSOPTIONcompsoc
  \usepackage[nocompress]{cite}
\else
  \usepackage{cite}
\fi

\ifCLASSINFOpdf
\else
\fi

\begin{document}

\title{Learning from the Good Ones: Risk Profiling-Based Defenses Against Evasion Attacks on DNNs}


\author{
\IEEEauthorblockN{Mohammed Elnawawy}
\IEEEauthorblockA{\textit{Department of Electrical and Computer Engineering} \\
\textit{University of British Columbia}\\
Vancouver, Canada \\
mnawawy@ece.ubc.ca}
\and
\IEEEauthorblockN{Gargi Mitra}
\IEEEauthorblockA{\textit{Department of Electrical and Computer Engineering} \\
\textit{University of British Columbia}\\
Vancouver, Canada \\
gargi@ece.ubc.ca}
\and
\IEEEauthorblockN{Shahrear Iqbal}
\IEEEauthorblockA{
\textit{National Research Council Canada}\\
Canada \\
shahrear.iqbal@nrc-cnrc.gc.ca}
\and
\IEEEauthorblockN{Karthik Pattabiraman}
\IEEEauthorblockA{\textit{Department of Electrical and Computer Engineering} \\
\textit{University of British Columbia}\\
Vancouver, Canada \\
karthikp@ece.ubc.ca}
}

\newcommand{\karthik}[1]{\todo[size=\small]{ #1}}
\newcommand{\gargi}[1]{\todo[size=\small]{ #1}}
\newcommand{\shahrear}[1]{\todo[size=\small]{ #1}}
\newcommand{\nawawytodo}[1]{\todo[size=\small]{ #1}}
\newcommand{\nawawy}[1]{{\color{blue} #1}}
\newcommand{\nawawyOld}[1]{{\color{black} #1}}

\maketitle
\begin{abstract}
Safety-critical applications such as healthcare and autonomous vehicles use deep neural networks (DNN) to make predictions and infer decisions. DNNs are susceptible to evasion attacks, where an adversary crafts a malicious data instance to trick the DNN into making \nawawyOld{wrong decisions} at inference time. \nawawyOld{Existing defenses \nawawyOld{that protect DNNs} against evasion attacks are either static or dynamic. Static defenses are computationally efficient but do not adapt to the evolving threat landscape, while dynamic defenses are adaptable but suffer from an increased computational overhead. To \nawawyOld{combine the best of both worlds,}
in this paper, we propose a novel \nawawyOld{risk profiling framework that uses a risk-aware strategy to}
selectively train static defenses using \nawawyOld{victim} instances that exhibit the most resilient features and are hence more resilient against an evasion attack. We hypothesize that training existing defenses on instances that are less vulnerable to the attack enhances the adversarial detection rate by reducing false negatives.}
\nawawyOld{We evaluate the efficacy of our risk-aware selective training strategy on a blood glucose management system} that demonstrates how training \nawawyOld{static anomaly detectors} indiscriminately may result in an increased false negative rate, which could be life-threatening \nawawyOld{in safety-critical applications}.
Our experiments show that selective training on the less vulnerable patients achieves a recall increase of up to 27.5\% with minimal impact on precision compared to indiscriminate training. 
\end{abstract}

\begin{IEEEkeywords}
risk profiling, evasion attacks, anomaly detectors, selective training, and blood glucose management.
\end{IEEEkeywords}
\section{Introduction}
\label{Section: Introduction}
Deep neural networks (DNNs) have gained traction in safety-critical applications \nawawyOld{such as} healthcare \cite{alanazi2022using, habehh2021machine, wiens2018machine, bhowmik2022deep, AI-Rad} \nawawyOld{and autonomous vehicles (AVs) \cite{bachute2021autonomous, chen2021exploring, xiao2022deep}}. 
However, DNNs are highly susceptible to adversarial attacks \cite{narodytska2017simple, fawaz2019adversarial, amini2024fast}\nawawyOld{, especially evasion attacks \cite{DBLP:journals/corr/KurakinGB16a, chernikova2019self, herath2021real}, which are prevalent since they are relatively easy to execute during deployment \cite{Boesch_2024, farinetti2018evasion}}. 
In evasion attacks, \nawawyOld{a DNN is tricked into misclassifying an adversarial sample at inference time, leading to poor accuracy \cite{szegedy2013intriguing, goodfellow2014explaining}}.
For example, \nawawyOld{an adversary may target \nawawyOld{DNN} models that predict blood glucose values to cause insulin overdose or underdose by manipulating patients' vital signs like previous blood glucose values or administered insulin dosage while ensuring the resulting glucose is within physiological limits to evade detection, leading to catastrophic consequences \cite{levy2022personalized}}.  

Researchers have proposed defense strategies to make DNNs resilient against evasion attacks \nawawyOld{including} adversarial training \cite{haroon2022adversarial, van2022defending, zhao2022adversarial}, training dataset strengthening \cite{xie2019feature, han2021evaluating, park2024adversarialfeaturealignmentbalancing, giraldo2020more}, model algorithm enhancement \cite{ghafouri2018adversarial, goodge2022lunar, 9833836, huang2022adversarial}, and anomaly detectors \cite{li2019mad, randhawa2024deep, qui2021strengthening, shu2022omni, elgarhy2023clustering, ahmed2022mitigating, wang2023evasion}. 
\nawawyOld{These defenses are either static or dynamic in nature. Static defenses are easier to implement, demonstrate higher accuracy on benign data, and are more computationally efficient. However, they cannot adapt to different attack strategies or the evolving behavior of victim instances \cite{wang2303adversarial}. Dynamic defenses, on the other hand, are more robust to evasion attacks because they adapt to evolving attack and victim behaviors. However, they suffer from degradation of benign data accuracy and high computational overhead at inference time. Therefore, they are not suitable for time-sensitive safety-critical applications
\cite{croce2022evaluating}. 

To bridge the gap between static and dynamic defenses, we propose a novel risk-aware selective training strategy that improves the adaptability of static defenses, while retaining their computational efficiency in the presence of an attack. Our risk-aware strategy \nawawyOld{is powered by} a risk profiling framework that selects training instances that show more resilience to the attack. \emph{The key idea is that instances that are less vulnerable to the evasion attack are usually a better representation of a typical distribution of benign data.} As a result, training static defenses to recognize a better distribution of benign data makes it easier for the defense technique to recognize malicious patterns generated by evasion attacks. 
}

In this work, we focus on training anomaly detectors for attack detection in safety-critical applications, e.g., \textit{k}NN, OneClassSVM, and MAD-GAN \cite{li2019mad}. 
\nawawyOld{The main issue with existing 
anomaly detectors is that they are often} indiscriminately trained on \nawawyOld{the entire dataset}
to capture the full spectrum of possible \nawawyOld{risk scenarios}
\cite{newaz2020adversarial, li2021defending, joe2021machine}. However, this strategy has three major problems. \nawawyOld{\textit{First}, it often yields} detectors that are less robust against evasion attacks due to the presence of noisy data samples which obscure learning meaningful patterns \nawawyOld{for malicious data detection} \cite{GUPTA2019466, shen2022adversarial}. \nawawyOld{\textit{Second}, it degrades the model’s generalizability, which is crucial for deploying models in diverse \nawawyOld{adversarial} settings \cite{alawad2019adversarial}. 
\textit{Third}, it incurs increased computational cost during training  \cite{he2020robustness}.} \nawawyOld{\emph{We hypothesize that training anomaly detectors using less vulnerable instances can improve malicious data detection by lowering the false negative rate.} We prioritize lower false negatives since higher false negatives in safety-critical applications may lead to deadly consequences whereas higher false positives may} \nawawyOld{lead to denial of service attacks 
and lack of availability, which are less severe in such systems.} 

Thus, our goal is to maximize the recall of existing anomaly detectors without causing much degradation to their precision. Towards this goal, we introduce a risk profiling framework that selectively trains existing \nawawyOld{anomaly detectors} on the most resilient instances to help them better differentiate between benign and malicious samples. This boosts the detection rate while overlooking noisy samples that impede the learning process. \nawawyOld{Our risk profiling framework consists of five steps. \textit{First}, it simulates the evasion attack. \textit{Second}, it quantifies the risk of manipulating data points at every point in time. \textit{Third}, it constructs a time-series risk profile for every victim. \textit{Fourth}, it groups risk profiles depending on their level of vulnerability to the attack. \textit{Fifth}, it uses instances that are less vulnerable to the attack to selectively train the anomaly detectors.}


\nawawyOld{We evaluate the efficacy of our proposed} \nawawyOld{risk profiling framework} 
on a blood glucose management system (BGMS) exposed to evasion attacks against Type-1 diabetes patients. \nawawyOld{In the context of a BGMS,} \nawawyOld{we define evasion attacks} as intentional glucose manipulations designed to deceive DNNs into predicting future glucose levels that result in an altered patient diagnosis.
We use the OhioT1DM dataset \cite{marling2020ohiot1dm} which includes physiological measurements of 12 Type-1 diabetes patients (six from 2018 and six from 2020). We also use a blood glucose prediction model from prior work~\cite{rubin2020deep}, which predicts future blood glucose values.

\nawawyOld{In summary, the contributions of this paper are \nawawyOld{twofold}: 
\begin{enumerate}
    \item A risk profiling framework to quantify the risk of an evasion attack \nawawyOld{on victims of safety-critical applications}
    and group them into different vulnerability clusters.
    \item A strategy to selectively train anomaly detectors on \nawawyOld{instances} with the most resilient features against the attack as identified by the risk profiling framework.
\end{enumerate}
}

\nawawyOld{The results of our experiments show that compared to indiscriminate training, selective training guided by our risk profiling framework achieves a recall increase of 27.5\% and 16.8\% on \textit{k}NN and OneClassSVM, respectively, with little to no impact on precision.
Furthermore, when trained on the less vulnerable patients, a MAD-GAN detector maintains a false negative rate of zero with no change to its precision, at a 75\% reduction in training set size as opposed to indiscriminately training it on the entire dataset.} \nawawyOld{Therefore, our risk profiling framework helps static anomaly detectors achieve lower false negatives with minimal impact on false positives.}

\section{Proposed Framework}
\label{Section:Proposed Framework}
In this section, we present \nawawyOld{our} risk profiling framework 
for selective training of existing \nawawyOld{anomaly detectors} to improve their detection capabilities. \nawawyOld{We rely on anomaly detectors that work in conjunction with the main DNN prediction model. The main DNN model remains unmodified since our proposed risk profiling framework is only used to train the anomaly detectors used to defend against adversarial attack samples.} 
The \nawawyOld{key idea of the proposed framework is to identify \nawawyOld{instances} that are \nawawyOld{more resilient} \nawawyOld{due to their natural physiology or driving habits}.
To do so, the proposed framework} categorizes victim instances into clusters of different risk levels depending on their vulnerability to \nawawyOld{the} evasion attack. \nawawyOld{Once it determines the most resilient \nawawyOld{instances}, the framework uses their past data to selectively train anomaly detectors to recognize the robust features that allow the instances to combat the evasion attack.}

\nawawyOld{Figure \ref{fig:framework} shows the proposed risk profiling framework, which consists of five steps. \nawawyOld{\textit{First}, the framework simulates the evasion attack by generating manipulated inputs to deceive the main DNN model and evaluate its vulnerabilities. \textit{Second}, it quantifies the amount of risk imposed on a victim by calculating the risk metrics at each time stamp to assess the impact of adversarial manipulations on individual data points. \textit{Third}, it constructs a continuous risk profile for each victim to capture their temporal patterns. The risk profile is a time-series representation of all risk values calculated in step 2. \textit{Fourth}, it uses unsupervised machine learning techniques to cluster time-series risk profiles into distinct risk categories, enabling differentiation of vulnerability levels. \textit{Fifth}, it incorporates clustering insights by selectively training anomaly detectors on the less vulnerable instances
to learn robust features that improve resilience against evasion attacks.}


\begin{figure}
    \centering
    \includegraphics[scale=0.25]{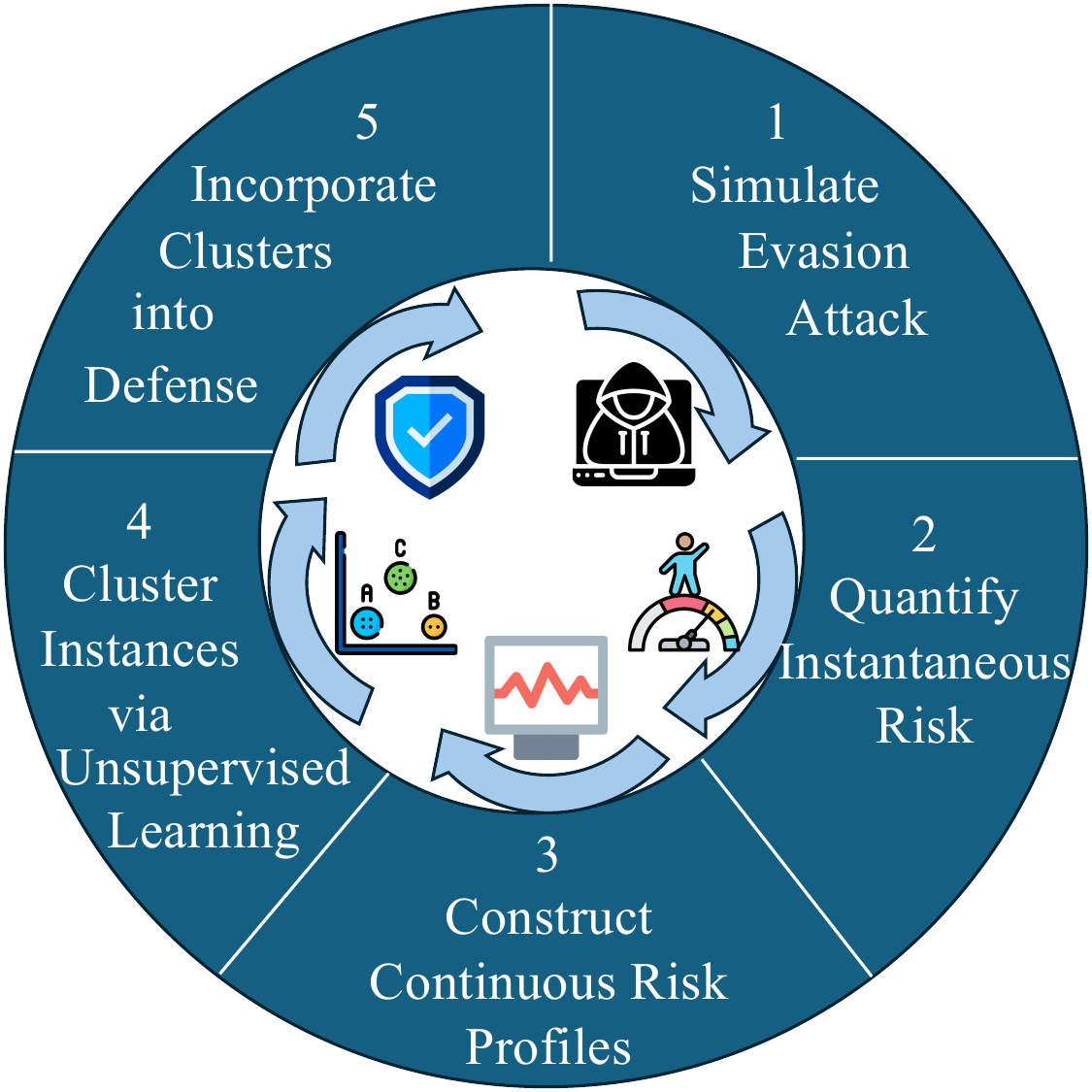}
    \caption[justification=centering]{The \nawawyOld{five steps of the} proposed risk profiling framework. }
    \label{fig:framework}
\end{figure}

}

\section{Blood Glucose Management System}
\label{Section:BGMS}
\begin{figure}
    \centering
    \includegraphics[scale=0.20]{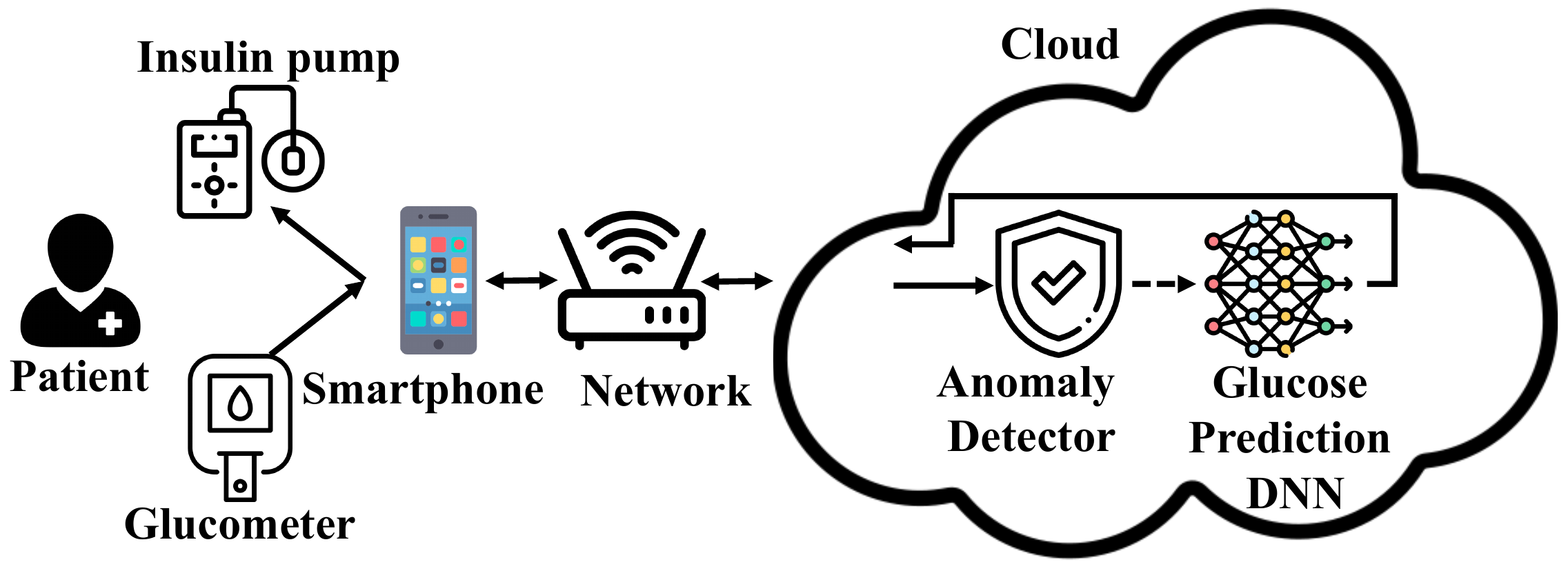}
    \caption[justification=centering]{A BGMS that uses a glucometer, insulin pump, DNN for insulin recommendations, and an anomaly detector to detect adversarial samples.  }
    \label{fig:BGMS}
\end{figure}

\nawawyOld{To evaluate the efficacy of our proposed risk profiling framework in enhancing the performance of existing anomaly detectors, we adopt and extend} the case study presented in Elnawawy et al.~\cite{elnawawy2024systematically} to simulate evasion attacks.
We consider \nawawyOld{a} BGMS \nawawyOld{(shown in Figure \ref{fig:BGMS})}
that consists of a continuous glucose monitor (CGM) that measures glucose at regular intervals and transmits it to a smart app running on a mobile device via Bluetooth. The app sends the measured glucose to \nawawyOld{the cloud, where an anomaly detector inspects glucose samples to flag any malicious patterns. If a glucose sample is deemed to be benign by the anomaly detector, it is used by the main} DNN \nawawyOld{model for processing and} \nawawyOld{future glucose predictions.} \nawawyOld{Next, the DNN model sends the predicted future glucose to the mobile app, which calculates the recommended insulin dose and enables the patient to}
approve it before the insulin pump infuses the corresponding insulin into \nawawyOld{his/her} body.

\nawawyOld{\textbf{Threat model.}} The attacker's \nawawyOld{\textit{goal}}
\nawawyOld{is to deceive the BGMS into mistakenly recommending an excessively high insulin dose which could lead the patient into a coma or even death.}
\nawawyOld{The attacker}\nawawyOld{'s \textit{strategy} is} to \nawawyOld{cause the glucose prediction DNN to} 
\nawawyOld{predict} a high future blood glucose level (hyperglycemia), when \nawawyOld{in reality} the patient has a low (hypoglycemia) or normal blood glucose. 
\nawawyOld{To do so, the adversary manipulates the victim's}
blood glucose levels to values \nawawyOld{that exceed} 125 mg/dL (hyperglycemic \nawawyOld{in a fasting state}) or 180 mg/dL (hyperglycemic \nawawyOld{two hours postprandial}).
\nawawyOld{We assume minimal \textit{capabilities} where the adversary} can only manipulate the CGM measurements \nawawyOld{by compromising} the Bluetooth stack via known exploits \cite{rasmussen2022blurtooth, Medtronic} to intercept and manipulate glucose measurements since many CGM devices 
use Bluetooth to transmit \nawawyOld{unencrypted glucose values} \cite{niu2025securing}. \nawawyOld{Manipulating other features remains beyond the attacker's capabilities. However, we assume the adversary can compromise the smartphone \cite{suarez2015compartmentation} to read these features and ensure the soundness of the generated adversarial samples.} 


\nawawyOld{\textbf{Target glucose model.}} Since the \nawawyOld{glucose prediction} algorithm used by smart apps is often confidential \cite{dreamed},
we approximated it using a time-series prediction model developed by Rubin-Falcone et al. \cite{rubin2020deep}, \nawawyOld{which
uses a bidirectional long short-term memory (LSTM) 
architecture.}
Rubine-Falcone et al. \cite{rubin2020deep} built two \nawawyOld{types of} models: (i) a personalized model for each patient trained on the patient's \nawawyOld{individual} data, and (ii) an aggregate model trained on the data of all patients. \nawawyOld{We use both types of models to simulate the evasion attack.}

\nawawyOld{\textbf{Dataset.}} To demonstrate the \nawawyOld{effect} of adversarial \nawawyOld{glucose values} on the target model\nawawyOld{'s predictions}, we use the OhioT1DM dataset \cite{marling2020ohiot1dm}, which was also used by the target model \cite{rubin2020deep} \nawawyOld{to evaluate} its accuracy. The dataset comprises physiological measurements of 12 Type-1 diabetes patients (six from 2018 and six from 2020). \nawawyOld{For the rest of this paper, we refer to the 2018 and 2020 patients as \textit{Subset A} and \textit{Subset B}, respectively.} The main features are the CGM measurements, finger-based measurements, basal insulin, bolus dose, carbohydrate intake, heart rate, sleeping patterns and acceleration, besides other physiological, and self-reported life-event features. The dataset spans eight weeks and consists of $\approx$10000 samples for training, and 2500 samples for testing, recorded at approximately five-minute intervals per patient.

\nawawyOld{\textbf{Attack algorithm.}} \nawawyOld{As for the evasion attack}, we use the universal robustness evaluation toolkit (URET), which is a general-purpose evasion attack framework for manipulating data points at inference time \cite{eykholturet}. To ensure that \nawawyOld{manipulated} CGM values respect physiological levels, we constrain them to be between 125 and 499 mg/dL for fasting \nawawyOld{scenarios}, since a hyperglycemic glucose level in a fasting state \nawawyOld{is greater than} 125 mg/dL, and between 180 and 499 mg/dL for postprandial \nawawyOld{scenarios}, since a hyperglycemic glucose level in a postprandial state \nawawyOld{is greater than} 180 mg/dL (499 mg/dL is the highest reported glucose level in the OhioT1DM dataset). 

\nawawyOld{\textbf{Anomaly detectors.} To test our framework, we use three anomaly detectors, \textit{k}NN, OneClassSVM, and MAD-GAN \cite{li2019mad}. We use \textit{k}NN, for its strength in handling sparse neighborhoods \cite{zhao2018review}, which better represent anomalies in medical data \cite{adler2015sparse, samariya2023detection, fang2015sparsity}, OneClassSVM for its strength in learning decision boundaries near benign data, making it effective for detecting rare or unusual patterns \cite{li2003improving}, and MAD-GAN for its strength in capturing multivariate time-series feature dependencies, which is well suited for safety-critical applications like healthcare and AVs \cite{li2019mad}.}
\section{Evaluation}
\label{Section: Evaluation}

\nawawyOld{In this section, we ask the following research questions:
\begin{enumerate}[label = {\textbf{RQ\arabic*:}}, align=left]
    \item Does indiscriminate training of anomaly detectors result in a higher false negative rate? \nawawyOld{If so, when?}
    \item What is the most suitable selective training strategy to prioritize lower false negatives in anomaly detectors?
\end{enumerate}
}

\nawawyOld{To answer the questions, we apply our proposed risk profiling framework to the BGMS discussed in Section \ref{Section:BGMS} and show how it can be used to enhance the performance of static anomaly detectors using selective training.}

\nawawyOld{\textbf{Step 1: Attack Simulation. }}\nawawyOld{In their demonstration of the URET evasion attack on the OhioT1DM dataset using the attack settings presented earlier, }\nawawyOld{Elnawawy et al.~\cite{elnawawy2024systematically} show that patients}
respond differently to the same attack \nawawyOld{settings} as they show different vulnerability levels \nawawyOld{to the attack. In particular, Elnawawy et al. \cite{elnawawy2024systematically} report attack success rates of mispredicting normal glucose as high glucose reaching up to 100.0\% while fasting, and 97.9\% postprandial on some patients of \textit{Subset B}, while others show success rates of only 67.4\% while fasting, and 44.2\% postprandial. This suggests that it is more challenging for URET to attack specific patients who show more resilience to the attack \cite{elnawawy2024systematically}.} \nawawyOld{We extended their experiments to test the URET attack on \textit{Subset A} (Appendix \ref{Appendix:Subset A Result}). The results confirm that different patients of \textit{Subset A} also show different vulnerability levels to the same evasion attack.}


\nawawyOld{\textbf{Steps 2 and 3: Risk Quantification. }}\nawawyOld{To quantify the instantaneous risk of an attack at every timestamp, \nawawyOld{our risk formula considers two factors:
    (1) magnitude of deviation, and 
    (2) severity/cost of deviation, between the benign and adversarial model predictions.}
The magnitude of deviation is essential for the risk formula since it determines \nawawyOld{the prediction's} state transition. 
For example, \nawawyOld{modifying} the blood glucose \nawawyOld{prediction} from 90 mg/dL to 210 mg/dL transitions a patient from a state of normal glucose to a state of hyperglycemic glucose. The severity of deviation is important since it weighs state transitions differently depending on the threats they pose to victim instances. For example, transitioning a diabetic patient from hypoglycemic to hyperglycemic glucose is more life-threatening than from normal to hyperglycemic glucose.}

\nawawyOld{In our case study}, we calculate the instantaneous risks of manipulating blood glucose values using Equation \ref{Eq: instantaneous error}:

\begin{equation}
    \label{Eq: instantaneous error}
    R_t = S * Z_t, \qquad t \subset \mathbb{N}
\end{equation}

\noindent where $R_t$ is the instantaneous risk at time unit $t$, $S$ is the severity/cost coefficient of mispredicting a patient's blood glucose level, and $Z_t$ is the difference in magnitude between the benign and adversarial glucose predictions at time unit $t$. $Z_t$ can be calculated using Equation \ref{Eq: Magnitude}:

\begin{equation}
    \label{Eq: Magnitude}
    Z_t = (y_t - f(x_t))^2
\end{equation}

\noindent where $y_t$ is the benign glucose prediction at time $t$, and $f(x_t)$ is the glucose prediction at time $t$ in the presence of an attack. The difference between $y_t$ and $f(x_t)$ is squared in Equation \ref{Eq: Magnitude} to weigh big errors more compared to small ones (inspired by the mean squared error) since larger glucose differences could lead to more serious conditions. \nawawyOld{Next, after the framework calculates instantaneous risk values, it combines them to generate a continuous time-series risk profile for every victim.}

Ideally, severity coefficients should be determined by specialists. However, we did not have access to such specialists. Hence, \nawawyOld{we used exponential coefficients since in healthcare contexts such as BGMS, state transitions (e.g., hypoglycemia to hyperglycemia) are inherently nonlinear in their impact on patient outcomes \cite{li2024analyzing, chan2010nonlinear, clevelandclinic_somogyi_effect}. Hence, exponential coefficients capture this nonlinearity by assigning disproportionately higher coefficients to more severe state transitions.}
Table \ref{tab:severity coefficients} shows an example of severity coefficients assigned to different state transitions. For instance, a severity coefficient of 64 is assigned to a \nawawyOld{diagnosis} of hyperglycemia when the actual state of the patient is supposedly hypoglycemic. Hypoglycemia to hyperglycemia \nawawyOld{misdiagnosis} is considered to be the worst case since the system would mistakenly predict an excessively high insulin dose, which could lead to fatal outcomes \cite{yale2018hypoglycemia, Hypoglycemia, DiLonardo_Altomara_2024}. 


\begin{table}
\centering
\caption{Severity Coefficients for Different State Transitions}
\begin{footnotesize}
\begin{tabular}{|l|l|l|}

\hline
\textbf{Benign} & \textbf{Adversarial} & \textbf{Severity Coefficient (S)} \\ \hline
Hypo            & Hyper                & 64                                \\ \hline
Normal          & Hyper                & 32                                \\ \hline
Hypo            & Normal               & 16                                \\ \hline
Hyper           & Hypo                 & 8                                 \\ \hline
Hyper           & Normal               & 4                                 \\ \hline
Normal          & Hypo                 & 2                                 \\ \hline

\end{tabular}
\end{footnotesize}
\label{tab:severity coefficients}
\end{table}

\nawawyOld{\textbf{Step 4: Clustering.}} \nawawyOld{Once the framework} 
generates patients' risk profiles, \nawawyOld{it uses} hierarchical clustering to identify less vulnerable and more vulnerable patients to the attack. \nawawyOld{In our case study,} we chose hierarchical clustering for three reasons \cite{Noble_2024}. \textit{First}, we do not need to specify the number of clusters in advance since it is difficult to know apriori. Instead, the resulting dendrogram can be pruned at the desired level according to the distances between clusters. \textit{Second}, the dendrogram helps to visually observe patients with similar physiological characteristics at different levels of the hierarchy. \textit{Third}, it is suitable for clinical research since it categorizes mixed populations into more homogeneous groups.

\begin{figure}
\centering
\includegraphics[width=0.45\textwidth]{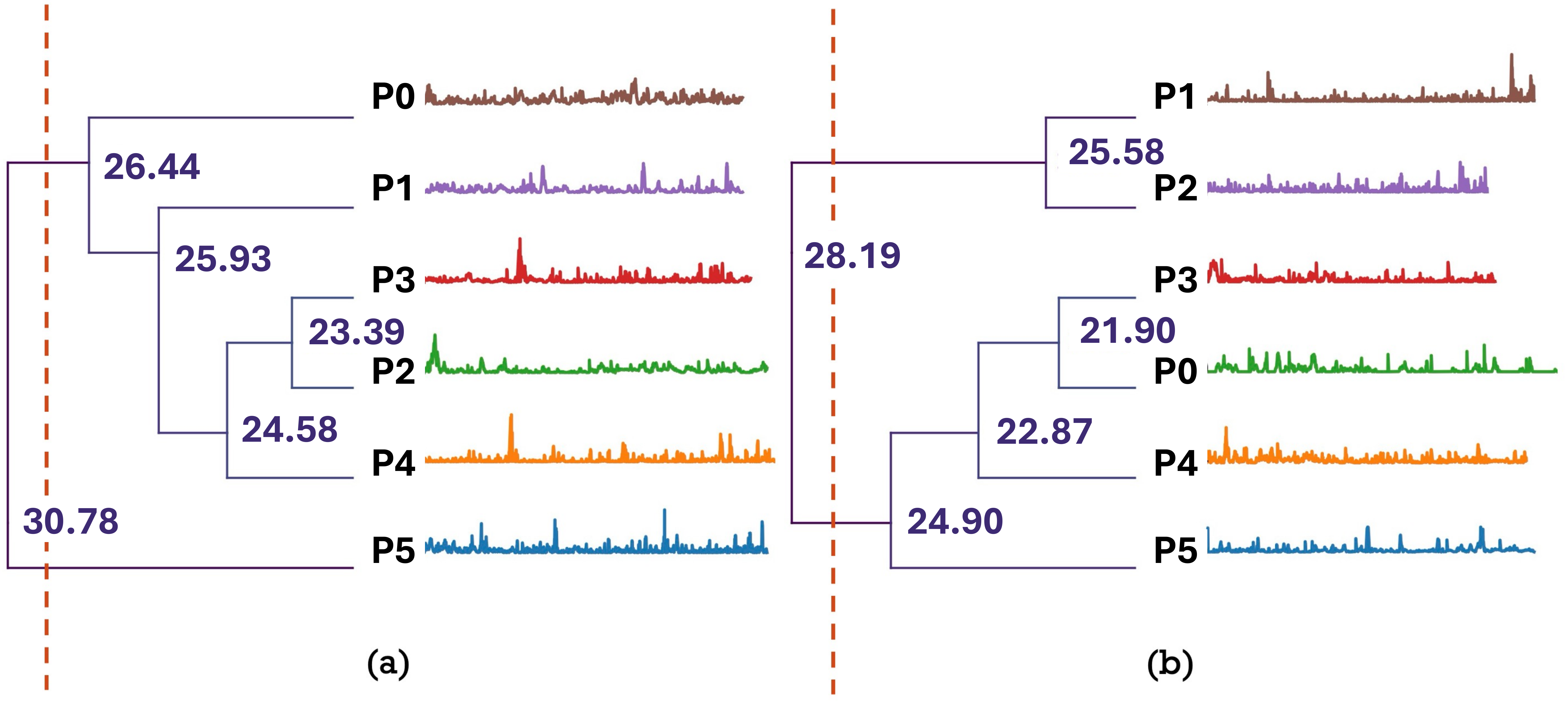}
\caption[Risk Profiling Clustering Results]{The results of hierarchically clustering the risk profiles from (a) \nawawyOld{\textit{Subset A}} and (b) \nawawyOld{\textit{Subset B}} of the OhioT1DM dataset. \nawawyOld{Based on the distance between the clusters, the dendrograms show that patients in either \textit{Subset} can be clustered into two groups - less and more vulnerable to the attack.}}
\label{fig:Clusters}
\end{figure}

Figure \ref{fig:Clusters} shows the time-series risk profiles for each of the six patients from (a) \nawawyOld{\textit{Subset A}} and (b) \nawawyOld{\textit{Subset B}}. It also shows the resulting dendrograms from hierarchically clustering the 12 patients. Based on the maximum distance between clusters in both cases, we decided to split the patients into two clusters: specifically, patients 0, 1, 2, 3, and 4 from \nawawyOld{\textit{Subset A}} belong to one cluster, and patient 5 belongs to the other cluster. Similarly, patients 0, 3, 4, and 5 from \nawawyOld{\textit{Subset B}} belong to one cluster, and patients 1 and 2 belong to another cluster. 
\nawawyOld{By cross-checking the resulting clusters with the misclassification percentages} \nawawyOld{due to the attack} \nawawyOld{reported in Elnawawy et al.~\cite{elnawawy2024systematically},} \nawawyOld{on \textit{Subset B} and our extended experiments on \textit{Subset A} (Appendix \ref{Appendix:Subset A Result}), }
we notice that patient 5 from \nawawyOld{\textit{Subset A}} and patients 1 and 2 from \nawawyOld{\textit{Subset B}} (placed in separate clusters by our risk profiling framework) tend to have the lowest misclassification percentage, meaning that these patients were \nawawyOld{less} vulnerable to the URET attack. On the other hand, the rest of the patients showed a relatively higher misclassification percentage, indicating that they were \nawawyOld{more} vulnerable to the attack. 
\nawawyOld{These observations enable us to  label the clusters according to patients' misclassification percentages as either less or more vulnerable to the URET attack.}
The obtained clusters are shown in Table \ref{tab:clusters}.

\begin{table}[]
\centering
\caption{Clusters of Patient Vulnerability to URET Attack}
\label{tab:clusters}
\begin{tabular}{|ll|lllll|}
\hline
\multicolumn{2}{|l|}{\textbf{Less Vulnerable}} & \multicolumn{5}{l|}{\textbf{More Vulnerable}} \\ \hline
\multicolumn{1}{|c|}{\textit{\textbf{Subset A}}} & \multicolumn{1}{c|}{\textit{\textbf{Subset B}}} & \multicolumn{3}{c|}{\textit{\textbf{Subset A}}} & \multicolumn{2}{c|}{\textit{\textbf{Subset B}}} \\ \hline
\multicolumn{1}{|l|}{\multirow{2}{*}{p5}} & p1 & \multicolumn{1}{l|}{p0} & \multicolumn{1}{l|}{p1} & \multicolumn{1}{l|}{p2} & \multicolumn{1}{l|}{p0} & p3 \\ \cline{2-7} 
\multicolumn{1}{|l|}{} & p2 & \multicolumn{1}{l|}{p3} & \multicolumn{2}{l|}{p4} & \multicolumn{1}{l|}{p4} & p5 \\ \hline
\end{tabular}
\end{table}

\begin{figure}
\centering
\includegraphics[width=0.3\textwidth]{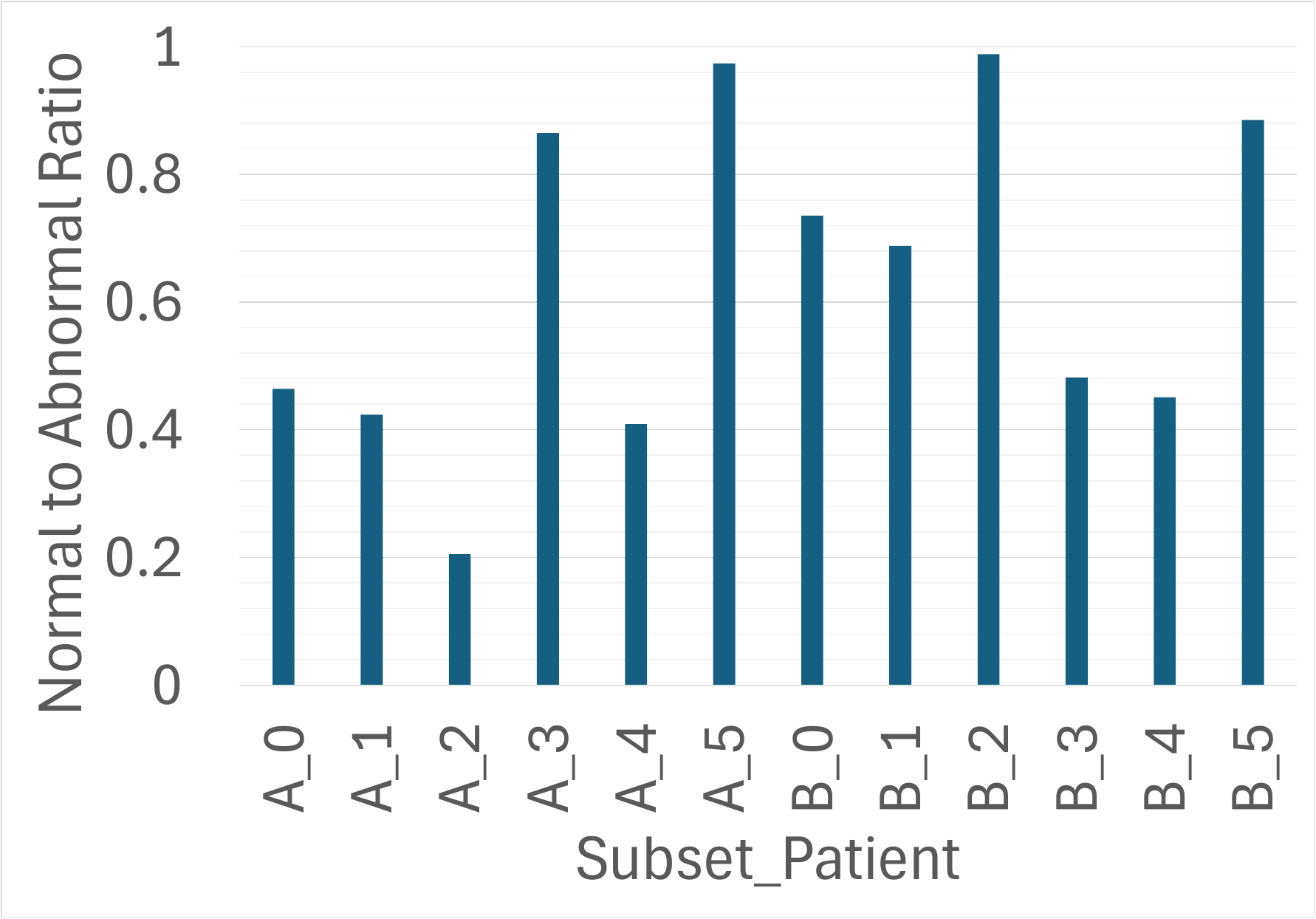}
\caption[Ratio of Normal to Abnormal Instances]{Ratio of normal to abnormal data instances in the benign trace of the patients. \nawawyOld{Less vulnerable patients tend to have higher ratios while more vulnerable patients tend to have lower ratios.}}
\label{fig:Ratio_Normal_To_Abnormal}
\end{figure}

To further analyze the obtained clusters, we plot the ratio of normal to abnormal (i.e., hypoglycemic or hyperglycemic) data points in the original benign trace of the 12 patients in Figure \ref{fig:Ratio_Normal_To_Abnormal}. We find that patient 5 from \nawawyOld{\textit{Subset A}} and patient 2 from \nawawyOld{\textit{Subset B}}\nawawyOld{, who belong to the less vulnerable cluster shown in Table \ref{tab:clusters},} show the highest \nawawyOld{benign} normal to abnormal glucose data points ratio. 
On the other hand, patient 2 from \nawawyOld{\textit{Subset A}} (\nawawyOld{more} vulnerable cluster) shows the lowest \nawawyOld{benign} normal to abnormal glucose data points ratio. 

\begin{figure}
\centering
\includegraphics[width=0.489\textwidth]{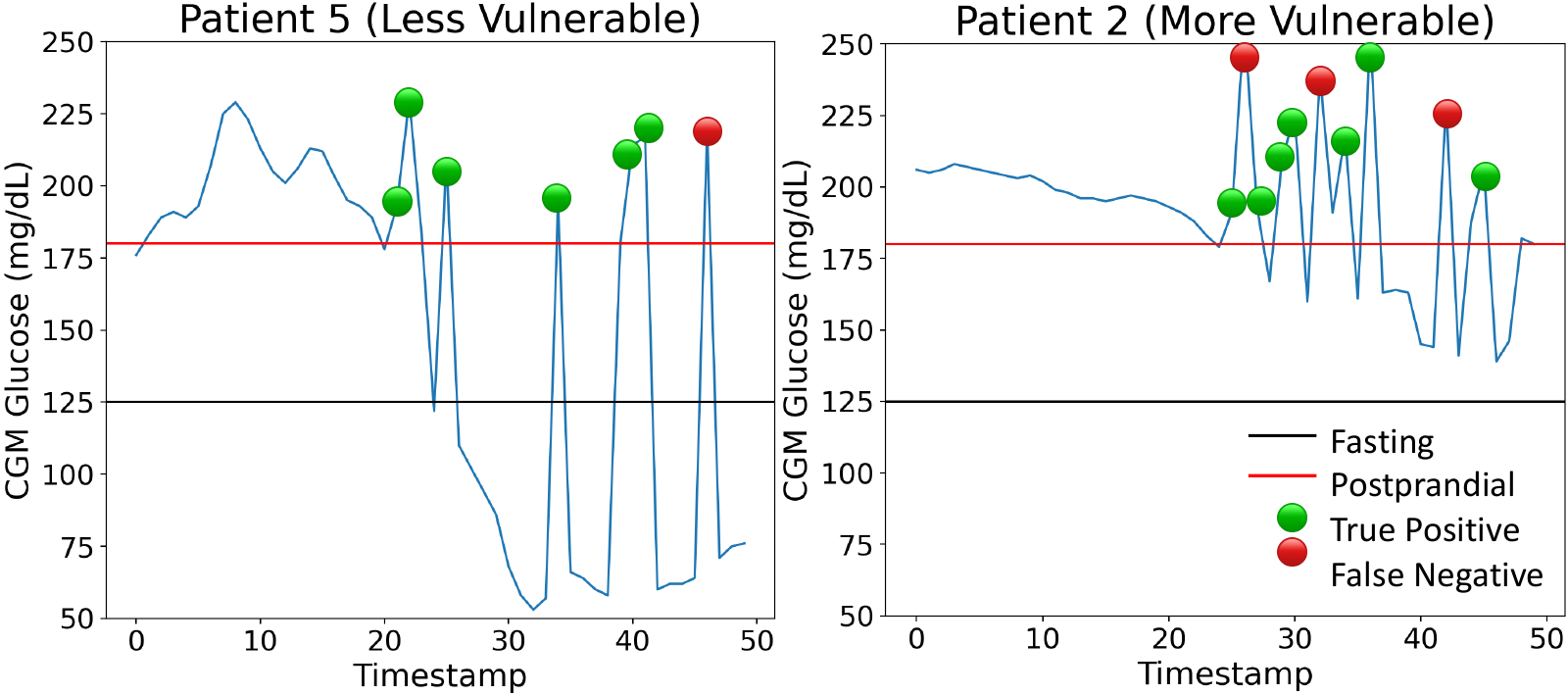}
\caption[Sample Glucose Traces with False Negatives and True Negatives]{\textit{k}NN anomaly detection on sample glucose traces of patients 5 and 2 from \nawawyOld{\textit{Subset A}. Indiscriminately training the detector yields a higher false negative rate on patient 2 (more vulnerable) than patient 5 (less vulnerable).}}
\label{fig:Plot_FN_TP_kNN}
\end{figure}

\nawawyOld{To demonstrate the issue of indiscriminate training \nawawyOld{\textbf{(RQ1)}}, we train a \textit{k}NN anomaly detector using data from all 12 patients of the OhioT1DM dataset. Figure \ref{fig:Plot_FN_TP_kNN} shows sample CGM glucose traces of patients 5 and 2 from \nawawyOld{\textit{Subset A}}. The black and red horizontal lines show the maximum normal glucose values in fasting (125 mg/dL) and postprandial (180 mg/dL) states, respectively. Green dots mark malicious glucose measurements that were successfully flagged by the anomaly detector (i.e., true positives), while red ones mark the missed malicious glucose measurements (i.e., false negatives). The figure shows that indiscriminately training the anomaly detector offers inequitable protection for the two patients since it flagged a higher percentage of adversarial samples from the \nawawyOld{less} vulnerable patient (i.e., patient 5) than the more vulnerable patient (i.e., patient 2). This indicates that the rate of false negatives is much higher for the \nawawyOld{more} vulnerable patient (patient 2) than for the less vulnerable patient (patient 5). }

\begin{figure}
\centering
\includegraphics[width=0.49\textwidth]{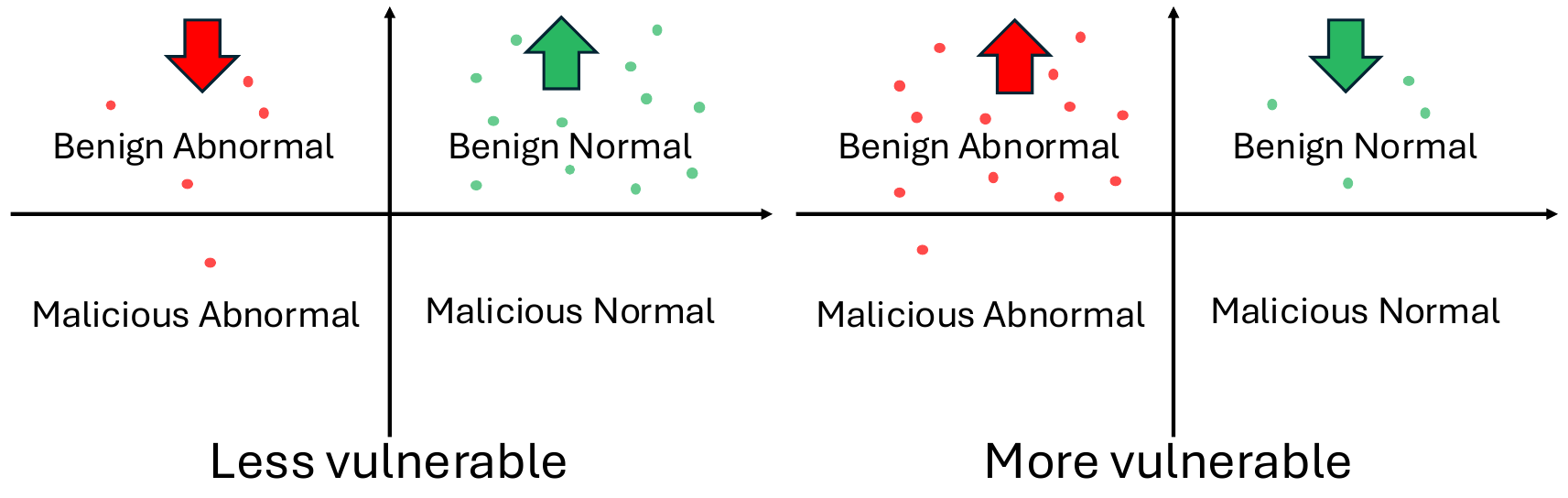}
\caption[Four Quadrants of Blood Glucose Samples]{\nawawyOld{The four quadrants of glucose samples: (a) benign normal: normal glucose in absence of attack, (b) benign abnormal: high or low glucose in absence of attack, (c) malicious abnormal: samples intentionally manipulated to fall in the high or low glucose ranges, and (d) malicious normal: samples intentionally manipulated to fall in the normal glucose range.}}
\label{fig:quadrants}
\end{figure}

\label{Paragraph:AnswerRQ1}
\nawawyOld{To explain the difference in false negatives \nawawyOld{caused by} indiscriminate training (\textbf{RQ1}), we \nawawyOld{analyze} Figure \ref{fig:Ratio_Normal_To_Abnormal}.} 
\nawawyOld{The more vulnerable patient (A\_2) \nawawyOld{shows a lower} ratio of benign normal to abnormal glucose levels}, \nawawyOld{indicating a higher prevalence of abnormal samples in their benign traces (Figure \ref{fig:quadrants}).}
\nawawyOld{Consequently,} when an anomaly detector \nawawyOld{encounters} a malicious abnormal \nawawyOld{sample, it is more likely to misclassify it as benign because it interprets the abnormality as part of the patient’s normal physiological variability rather than a result of an attack, leading to an increased false negative rate.}
\nawawyOld{In contrast}\nawawyOld{, the less vulnerable patient (A\_5) has a higher ratio of benign normal to abnormal glucose} samples, so when the detector sees a malicious abnormal sample there is a higher chance of flagging the sample as malicious \nawawyOld{(Figure \ref{fig:quadrants}). Training on such patients allows the detector to better distinguish between benign and malicious abnormalities, reducing false negatives, albeit at the cost of more false positives (potentially). }

\nawawyOld{\textbf{Step 5: Anomaly Detector Enhancement (RQ2).}} Training defenses on the \nawawyOld{less} vulnerable patients has three benefits. \textit{First}, it helps learn the most robust features against the attack since the defense focuses on the most resilient, less attack-prone, and more generalizable data features, which helps drive false negatives down. \textit{Second}, it avoids the risk of overfitting to adversarial samples if trained on the \nawawyOld{more} vulnerable patients. This is because when trained on the \nawawyOld{more} vulnerable patients, the model becomes more sensitive to adversarial features, and overfits to specific attack patterns, reducing its generalization ability and driving false positives up. \textit{Third}, the \nawawyOld{less} vulnerable instances better represent a typical distribution of benign data \nawawyOld{as shown by the higher ratio of benign normal to abnormal glucose samples in Figure \ref{fig:Ratio_Normal_To_Abnormal}}, providing a more balanced \nawawyOld{strategy} that preserves benign accuracy while detecting attacks. Therefore, we hypothesize that
\emph{training defenses on patients \nawawyOld{who} are \textit{less} vulnerable to the attack \nawawyOld{enhances} model resilience by \nawawyOld{reducing} false negative rate.} 

\nawawyOld{We use four subsets of the OhioT1DM data to train the three anomaly detectors, \textit{k}NN, OneClassSVM, and MAD-GAN \cite{li2019mad}} (check Appendix \ref{Appendix:Anomaly Detectors} for model parameters). The ``\nawawyOld{Less} Vulnerable'' and ``\nawawyOld{More} Vulnerable'' subsets comprise patients shown in Table \ref{tab:clusters}. The ``Random Samples'' subset consists of three patients drawn at random, repeated for 10 different runs, and averaged to reduce random errors and improve the accuracy of the results. \nawawyOld{Since the less vulnerable subset is used to test our hypothesis of selectively training anomaly detectors,} we randomly sampled three patients in each run of the ``Random Samples'' experiments to test whether the improvement in \nawawyOld{less} vulnerable training (which included exactly three patients) was purely due to chance. Finally, \nawawyOld{to show the improvement of selective training on the less vulnerable instances,} we \nawawyOld{indiscriminately} train the three defenses on ``All Patients'' to evaluate the efficacy of our selective training strategy. \nawawyOld{We consider ``All Patients'' and ``Random Samples'' to be our baseline training strategies since they train the anomaly detectors in the absence of insights from our risk profiling framework.}

\begin{figure}
\centering
\includegraphics[width=0.32\textwidth]{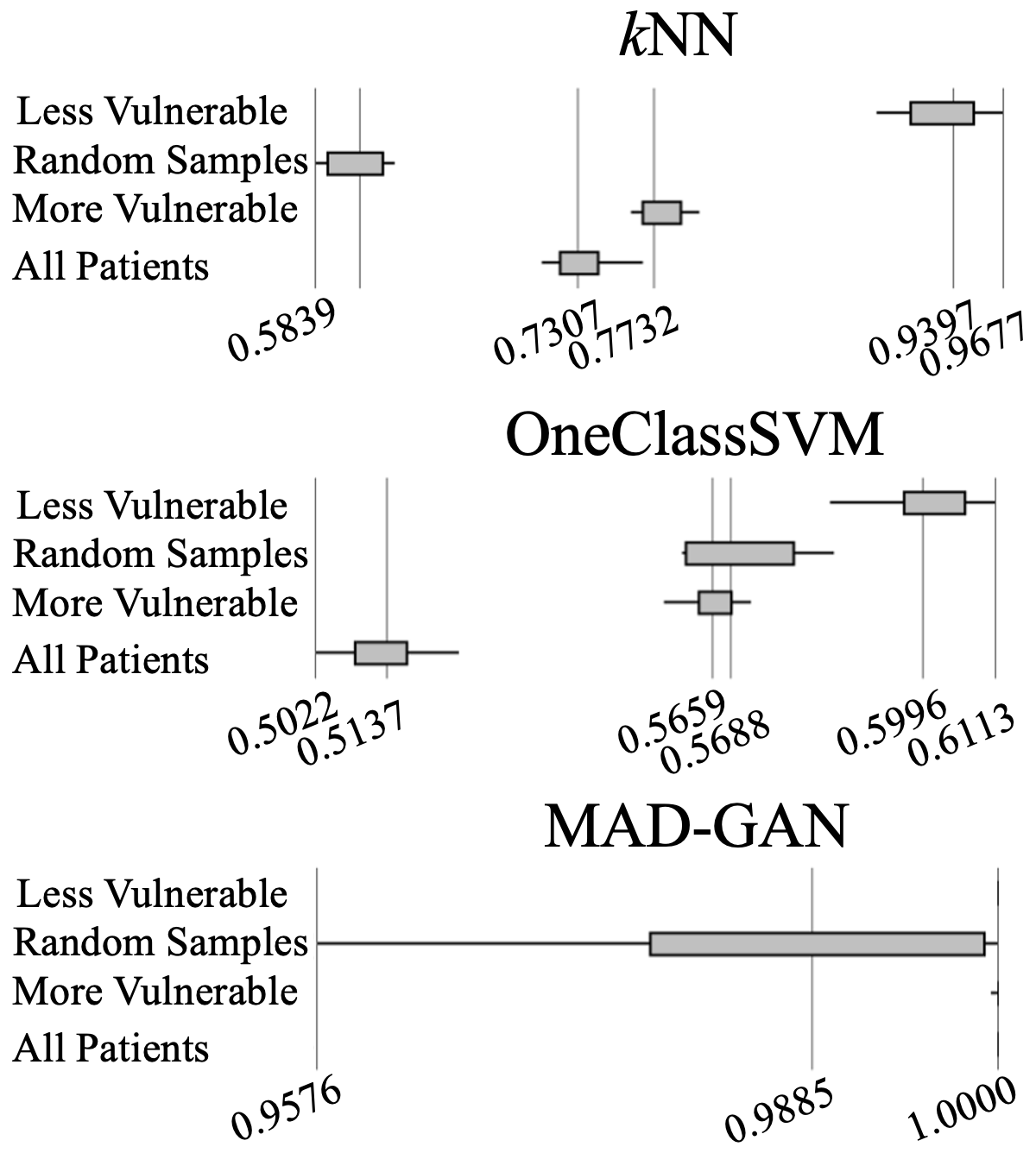}
\caption[Recall Results]{Recall results using \nawawyOld{\textit{k}NN, OneClassSVM, and MAD-GAN. Less vulnerable training achieves a recall increase of 27.5\% (\textit{k}NN), and 16.8\% (OneClassSVM) over indiscriminate training.}}
\label{fig:Recall}
\end{figure}

\nawawyOld{We first consider the recall of the detectors under selective training.} Figure \ref{fig:Recall} shows the results of the recall achieved for each of the training sets. \emph{We observe that training the three defenses on the \nawawyOld{less} vulnerable patients achieves the highest recall among all the subsets, for all three detectors.} 
In the case of \textit{k}NN and OneClassSVM, training using the \nawawyOld{less} vulnerable patients shows a significant improvement compared to \nawawyOld{indiscriminately} training using \nawawyOld{the entire dataset, achieving a percentage increase of 27.5\% and 16.8\% on \textit{k}NN and OneClassSVM, respectively. Moreover, the recall of the \nawawyOld{less} vulnerable training surpasses that of the} \nawawyOld{more} vulnerable or randomly sampled patients.
In the case of MAD-GAN, training on the \nawawyOld{less} vulnerable patients achieves the same recall as training on \nawawyOld{the entire dataset} (recall of 1), albeit at a \nawawyOld{75\%} reduction of training set size, \nawawyOld{ensuring better scalability with large, high-dimensional, non-linear, or complex datasets. 
}

\begin{figure}
\centering
\includegraphics[width=0.32\textwidth]{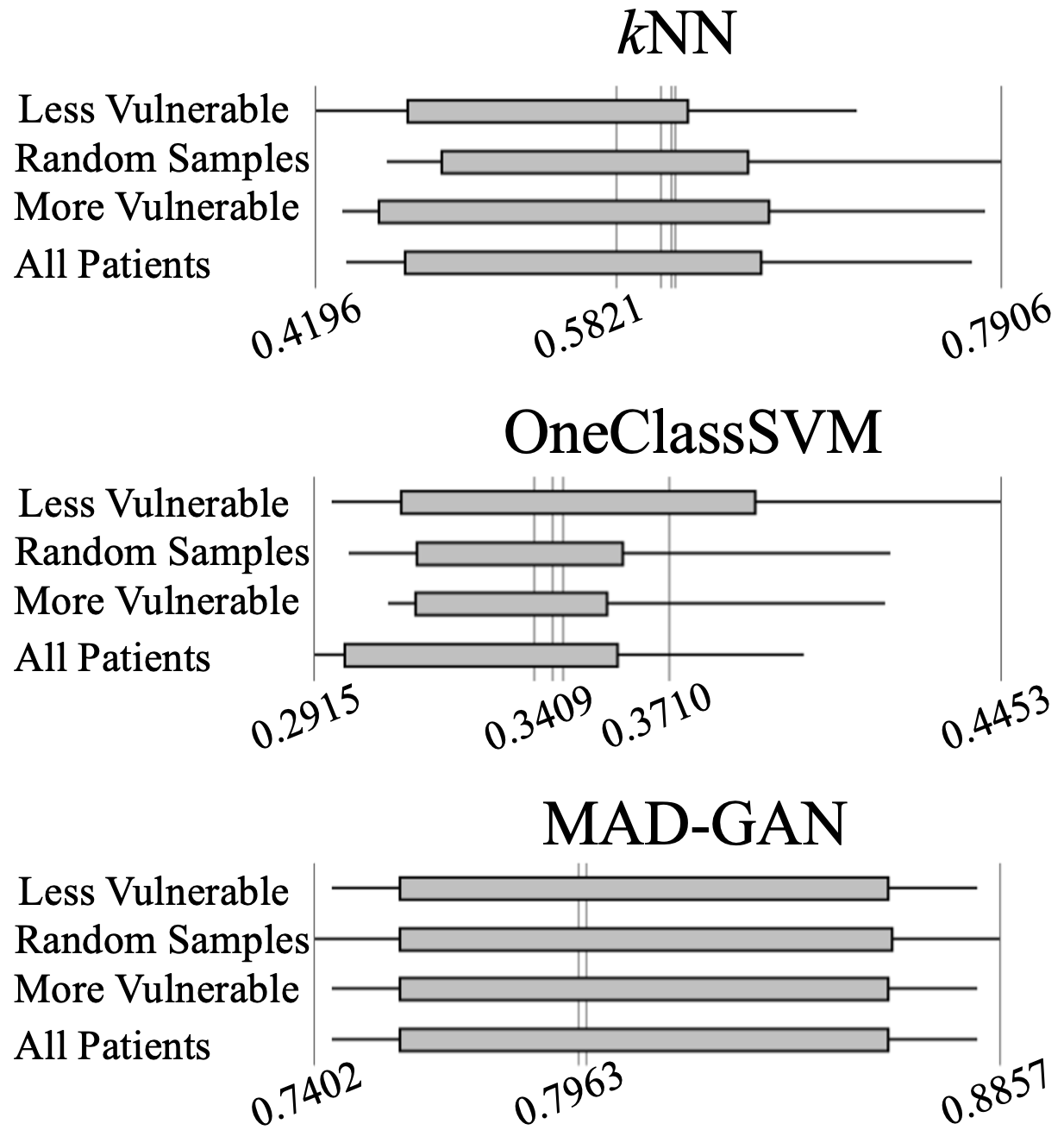}
\caption[Precision Results]{Precision results using \nawawyOld{\textit{k}NN, OneClassSVM, and MAD-GAN. Less vulnerable training yields a precision drop of 5\% (\textit{k}NN), and an increase of 7.5\% (OneClassSVM) over indiscriminate training.}}
\label{fig:Precision}
\end{figure}

We consider the detectors' precisions under selective training.
The precision results of \textit{k}NN shown in Figure \ref{fig:Precision} \nawawyOld{demonstrate} the trade-off between false negatives and false positives since an increase in \textit{k}NN's recall comes at the expense of \nawawyOld{a 5\% reduction in} precision with the \nawawyOld{less} vulnerable training. On the other hand, OneClassSVM shows \nawawyOld{a 7.5\% increase in precision when trained} using the \nawawyOld{less} vulnerable patients. This may be attributed to \textit{k}NN's sensitivity to the data distribution since glucose data is non-uniformly distributed or has varying densities in different regions;  hence \textit{k}NN may label sparse points as anomalies, despite being valid. Conversely, OneClassSVM is less sensitive to density variations since it creates a global model leading to lower false positives. As for MAD-GAN, all training subsets achieved similar precision. Thus, 
\nawawyOld{\textit{k}NN and MAD-GAN} suffered a small to no loss in precision\nawawyOld{, while OneClassSVM's precision improved under selective training.}  

To further investigate the \nawawyOld{recall-precision trade-off, we calculate their harmonic mean (i.e., F1-score)} \nawawyOld{(Appendix \ref{Appendix:F1-score Results})}.
We notice that \nawawyOld{selective} training on the \nawawyOld{less} vulnerable \nawawyOld{patients} significantly improves the performance of \textit{k}NN \nawawyOld{with an F1-score increase of 7.3\% compared to indiscriminate training despite the 5\% reduction in precision. This indicates that the combined effect of recall and precision captured by their harmonic mean shows an increase in anomaly detection performance, which highlights the efficacy of the proposed framework. On the other hand, OneClassSVM shows an F1-score increase of 10.9\% compared to indiscriminate training. The results show that despite potential increases in the false positive rate resulting from the recall-precision trade-off, selective training offers an improvement in the combined adversarial detection rate. }

\section{Limitations and Future Work}
\label{Section: Limitations and Future Work}
Our framework has \nawawyOld{four} main limitations. 
\textit{First}, it assumes 
training and testing data are drawn from the same distribution, 
which does not consider concept drifts \cite{zhang2022towards}. This 
leads to a failure to generalize to different data distributions and a failure to adapt to varying environments. For example, a risk profiler trained on senior patients' data may fail with young ones.
\textit{Second}, we use offline training to build a static risk profiler, which does not consider potential future dataset shifts. For example, patients move from high-risk to low-risk categories as they recover from medical conditions after the risk profiler has already classified them. \nawawyOld{\textit{Third}, we used} \nawawyOld{a single case study and a single attack algorithm}
\nawawyOld{to test the efficacy of our proposed framework. More datasets and} \nawawyOld{algorithms} \nawawyOld{are needed for a more thorough evaluation.}
In the future, we plan to build a risk profiler that uses online learning to consider varying attack environments, different attack algorithms, and potential dataset shifts to design a more \nawawyOld{adaptive} defense. 
\nawawyOld{\textit{Fourth}, our choice of severity coefficients is a direct threat to validity since it may impact the correctness of the risk profiles. In the future, we plan to conduct a sensitivity analysis on coefficient choice to further study this problem.}
\section{Conclusion}
\label{Section: Conclusion}

In this paper, we propose a risk profiling framework that \nawawyOld{bridges the computational overhead gap between static and dynamic defenses against evasion attacks. The proposed framework enhances the adaptability of static defenses to various threat levels using a novel risk-aware selective training strategy that} improves adversarial detection rate.
The framework generates time-series risk profiles \nawawyOld{for every victim} and \nawawyOld{clusters} \nawawyOld{them} into different risk categories based on their vulnerabilities to evasion attacks.
We show that selectively training \nawawyOld{static} anomaly detectors on 
\nawawyOld{the less vulnerable victims}
enhances \nawawyOld{their} detection rates.
\nawawyOld{We evaluated}
the proposed framework on a Type-1 diabetes case study.
Our results show \nawawyOld{that selective training surpasses indiscriminate training with} a reduction in false negatives across three \nawawyOld{anomaly detectors} (\textit{k}NN, OneClassSVM, and MAD-GAN),
\nawawyOld{achieving a recall increase of up to 27.5\% with} minimal impact on false positives. 

\section*{Acknowledgements}
This project is supported by collaborative research funding from the National Research Council of Canada’s Digital Health and Geospatial Analytics Program, UBC Four-Year Fellowships (FYF) and the Natural Sciences and Engineering Research Council of Canada (NSERC).

\bibliographystyle{unsrt} 

\appendices
\section{Subset A Results}
\label{Appendix:Subset A Result}

\begin{figure}[H]
\centering
\includegraphics[width=0.39\textwidth]{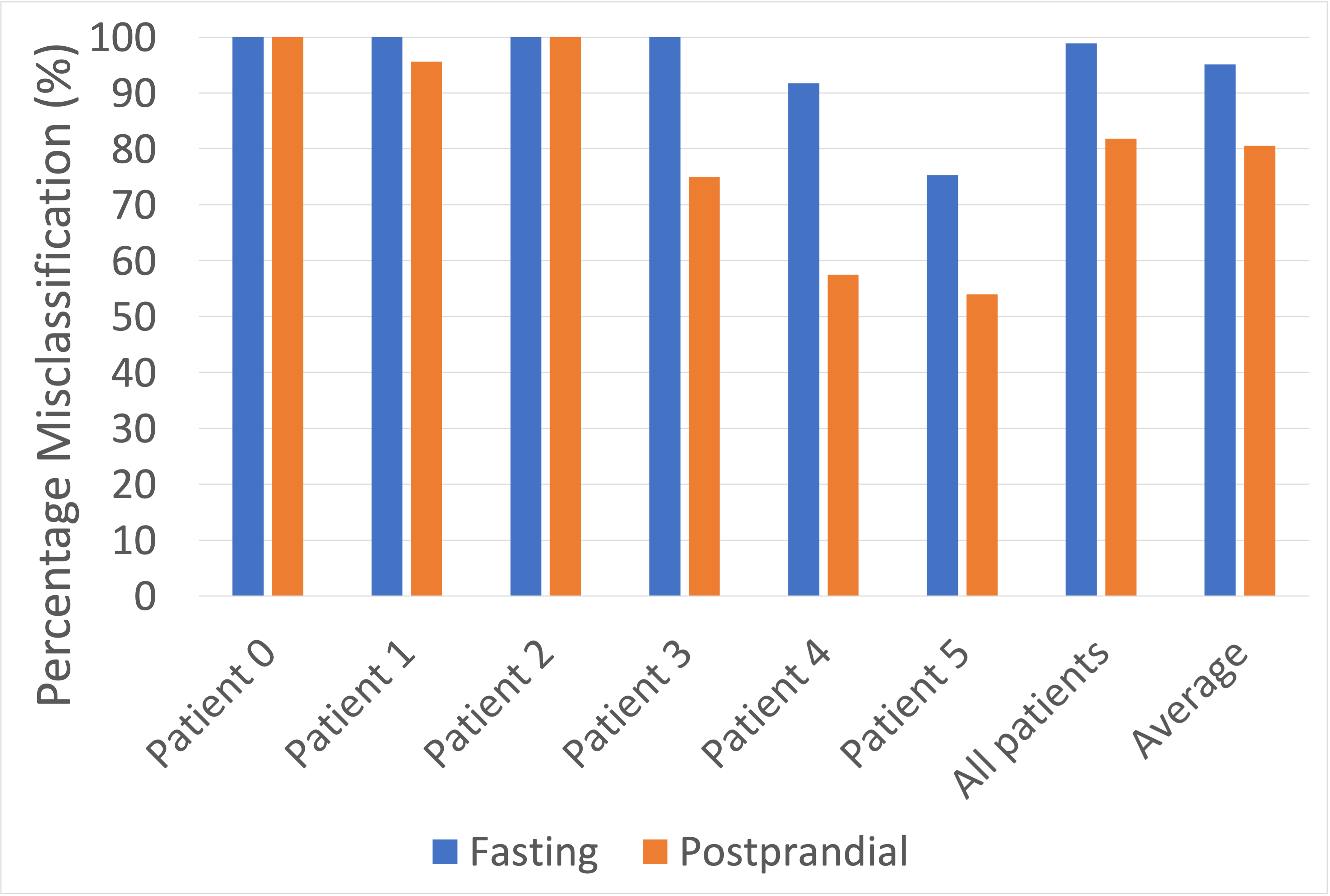}
\caption[Percentage of Normal Glucose Mispredictions - \textit{Subset A}]{Percentage of originally normal glucose instances that are misdiagnosed as hyperglycemic. ``Patient \textit{i}'' shows the results of the personalized model for the $\textit{i}^{th}$ patient, ``All patients'' shows the results of the aggregate model trained on the data of all patients, and ``Average'' shows the average results of the 7 models.}
\label{fig:2018_Normal_to_Hyper}
\end{figure}

\begin{figure}[H]
\centering
\includegraphics[width=0.39\textwidth]{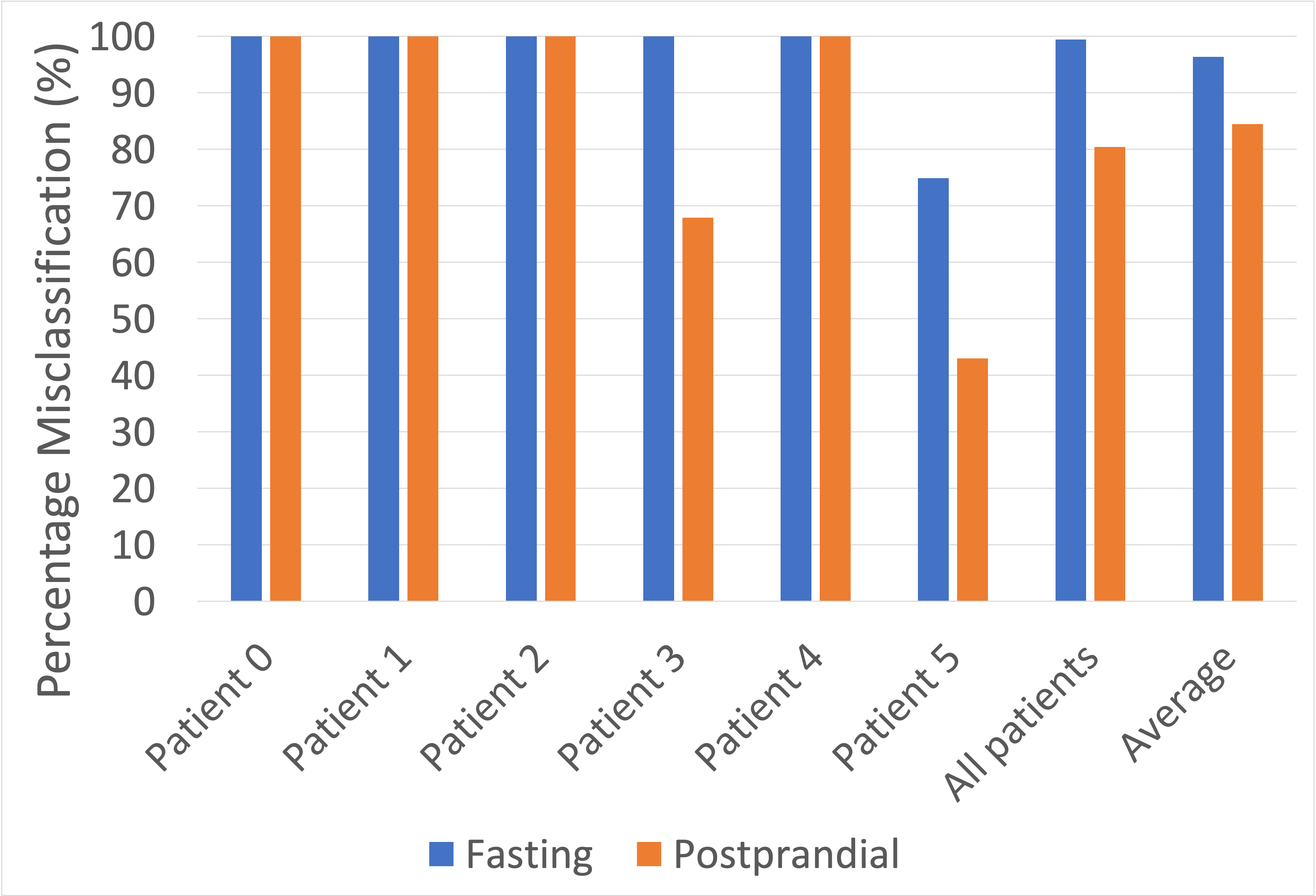}
\caption[Percentage of Hypoglycemic Glucose Mispredictions - \textit{Subset A}]{Percentage of originally hypoglycemic glucose instances that are misdiagnosed as hyperglycemic. ``Patient \textit{i}'' shows the results of the personalized model for the $\textit{i}^{th}$ patient, ``All patients'' shows the results of the aggregate model trained on the data of all patients, and ``Average'' shows the average results of the 7 models.}
\label{fig:2018_Hypo_to_Hyper}
\end{figure}
\section{Anomaly Detectors}
\label{Appendix:Anomaly Detectors}
In this appendix we elaborate on the anomaly detectors used to test our risk profiling framework.

\textbf{\textit{k}NN.} We use the KNeighborsClassifier implementation of the scikit-learn Python library with the following model parameters:
\begin{itemize}
    \item Number of neighbors = 7
    \item Weights = uniform
    \item Algorithm = auto
    \item Leaf size = 30
    \item p = 2
    \item Metric = minkowski
    \item Metric params = None
\end{itemize}

\textbf{OneClassSVM.} We use the OneClassSVM implementation of the scikit-learn Python library with the following model parameters:
\begin{itemize}
    \item Kernel = sigmoid
    \item Degree = 3
    \item Gamma = auto
    \item Coef0 = 10
    \item Tol = 0.001
    \item Nu = 0.5
    \item Shrinking = True
    \item Cache size = 200
    \item Max iter = -1
\end{itemize}

\textbf{MAD-GAN \cite{li2019mad}.} MAD-GAN is an unsupervised anomaly detection technique for multivariate time-series data. It uses a generative adversarial network (GAN) with long short-term memory recurrent neural networks (LSTM-RNN) as the generator and discriminator. MAD-GAN captures temporal correlations and latent interactions among features to detect anomalies using a novel anomaly score called discrimination and reconstruction anomaly score (DR-Score). We use the following model parameters in our adoption of MAD-GAN:
\begin{itemize}
    \item Number of epochs = 100
    \item Number of signals = 4
    \item Number of generated features = 4
    \item Sequence length = 12
    \item Sequence step = 1
\end{itemize}
\section{F1-score Results}
\label{Appendix:F1-score Results}
\begin{figure}[H]
\centering
\includegraphics[width=0.32\textwidth]{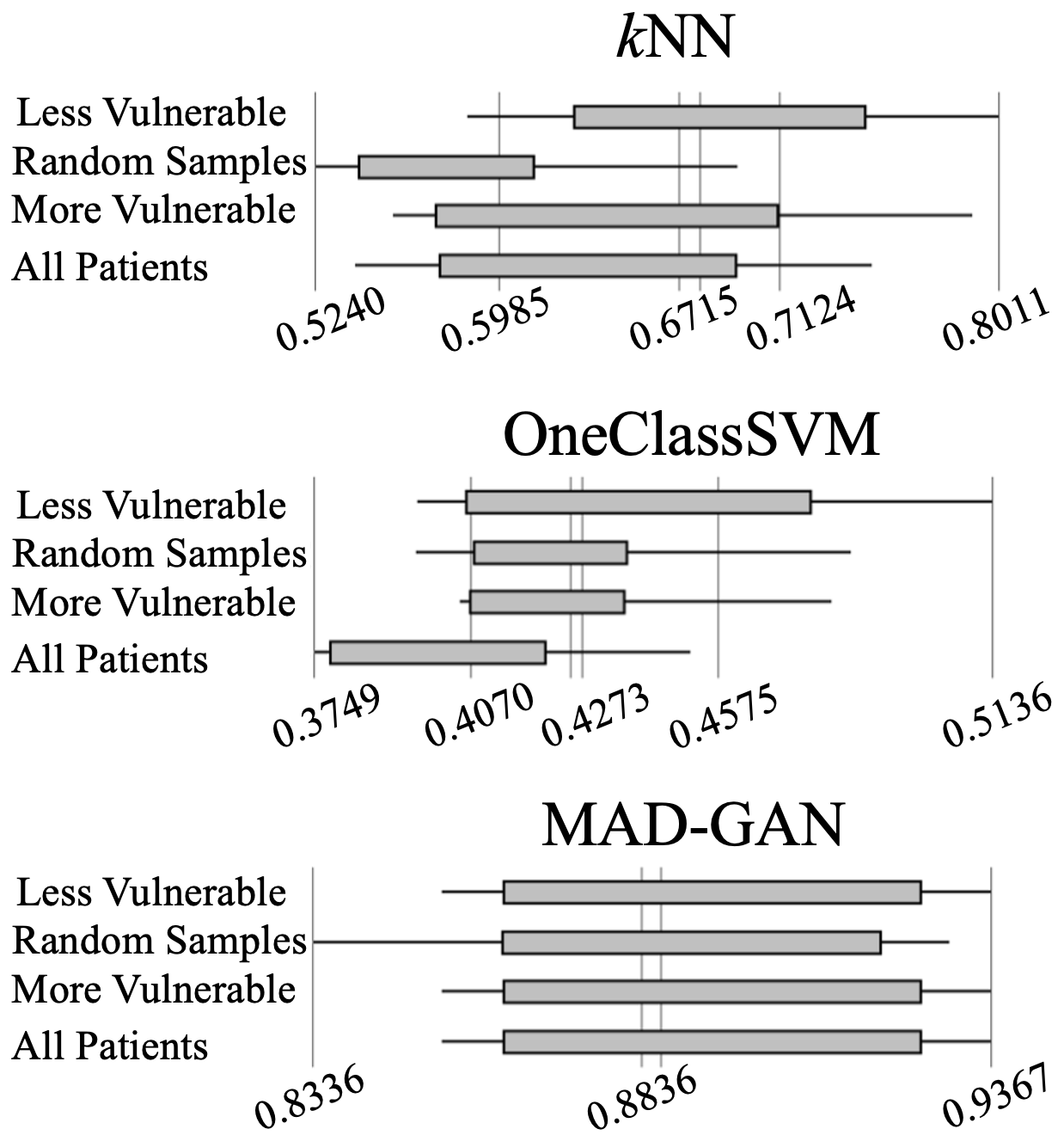}
\caption[F1-score Results]{F1-score results using \nawawyOld{\textit{k}NN, OneClassSVM, and MAD-GAN. Less vulnerable training achieves an F1-score increase of 7.3\% (\textit{k}NN), and 10.9\% (OneClassSVM) over indiscriminate training.} }
\label{fig:F1-score}
\end{figure}
\nawawyOld{
\section{Discussion}
In our experiments, we trained anomaly detectors on the less vulnerable patients and independently tested the entire set of patients and then averaged the results to aggregate them in box plots. That means that we conducted experiments where the test set consisted of only the more vulnerable patients that were not seen during the training stage. The resulting anomaly detection rates were similar to those obtained by testing on the less vulnerable patients. This demonstrates our framework's resilience to overfitting due to training only on the less vulnerable patients when tested on the OhioT1DM dataset. Nevertheless, we acknowledge that more rigorous testing on different datasets and attack algorithms is needed before confidently claiming the framework's resilience to overfitting and its generalizability to other domains. For this reason, we plan to extend our experiments to other healthcare datasets, the domain of autonomous vehicles (AVs), and other attack algorithms in our next publications. To further validate our work, we believe that our proposed framework could benefit from a mathematical model formulation to capture its full dynamics.

The proposed framework addresses concept drift through an iterative process that regularly reassesses patient risk profiles and continuously updates them as new data become available to strengthen defenses against the evolving threat landscape. As patient conditions evolve, so do their risk levels: those showing increased resilience against adversarial attack are incorporated into the retraining process, while those becoming more vulnerable are excluded from the occasional retraining. This continuous refinement ensures that the defense adapts over time to maintain robustness without sacrificing accuracy.

}

\IEEEpeerreviewmaketitle

\end{document}